\documentstyle[amssymb,prb,aps,psfig,epsf]{revtex}

\begin{document}
\title{Local density of states in superconductor-strong ferromagnet structures}
\author{F.S.Bergeret $^{1 }$, A.F. Volkov$^{1,2}$ and K.B.Efetov$^{1,3}$}
\address{$^{(1)}$Theoretische Physik III,\\
Ruhr-Universit\"{a}t Bochum, D-44780 Bochum, Germany\\
$^{(2)}$Institute of Radioengineering and Electronics of the Russian Academy%
\\
of Sciences, 103907 Moscow, Russia \\
$^{(3)}$L.D. Landau Institute for Theoretical Physics, 117940 Moscow, Russia }
\maketitle

\begin{abstract}
 We study the dependence of the local density of states (LDOS) on coordinates for
a superconductor-ferromagnet (S/F) bilayer and a S/F/S structure assuming
that the exchange energy $h$ in the ferromagnet is sufficiently large: $%
h\tau >>1,$ where $\tau $ is the elastic relaxation time. This limit cannot
be described by the Usadel equation and we solve the more general Eilenberger
equation. We demonstrate that, in the main approximation in the parameter $%
(h\tau )^{-1}$, the proximity effect does not lead to a modification of the
LDOS in the S/F system and a non-trivial dependence on coordinates shows up
in next orders in $\left( h\tau \right) ^{-1}.$ In the S/F/S sandwich the
correction to the LDOS is nonzero in the main approximation and depends on the phase difference
between the superconductors. We also calculate the superconducting critical
temperature $T_{c}$ for the bilayered system and show that it does not
depend on the exchange energy of the ferromagnet in the limit of large $h$
and a thick F layer.
\end{abstract}

\bigskip

\section{Introduction}

In recent years the interest in the proximity effect between superconducting
(S) and ferromagnetic (F) layers has increased considerably. The actual
progress in the preparation of high quality metallic multilayer systems
allows a careful study of the mutual interaction of superconductivity and
ferromagnetism in hybrid S/F\ structures. The proximity effect manifests
itself in {\it e.g.} changes of the density of states\cite{kontos} (DOS) or
in the dependence of the superconducting critical temperature $T_{c}$ on the
thickness of the F-layer (see e.g. Refs. \cite{koorevar,strunk,jiang}).

Many theoretical works have been devoted to the study of such structures, 
 which is not a simple task. In most of them the DOS and the critical
temperature of the superconducting transition $T_{c}$ were analyzed in the
``dirty limit''. This means that the retarded and advanced Green's
functions, which determine the thermodynamical properties, were obtained from
the Usadel equation (e.g Refs. \cite
{radovic,buzdin1,aarts,fazio,lazar,buzdin})  which is simpler than the
  more general Eilenberger equation. In other words, it was assumed that the mean free path $l$
is much shorter than any characteristic length (with exception of the Fermi
wave length), and that all energies involved in the problem are smaller
than $\tau ^{-1}$, where $\tau $ is the elastic relaxation time. In
particular, the condition $h\tau \ll 1$  has to be satisfied ($h$ is
the exchange energy). It was shown that in this limit the DOS in the F layer
oscillates and decays with increasing  exchange energy $h$. In the limit $%
h\tau \ll 1,$ the period of the oscillations and the decay length are
comparable to each other and are of the order $\sqrt{D/h}$. 

 Calculations in the opposite (purely ballistic) case were performed in
Refs.\cite{nazarov,minnesota}. Haltermanns and Valls \cite{minnesota} solved
the Bogolyubov and self-consistency equations numerically and presented the
results for the LDOS in a F/S structure. Zareyan and co-workers \cite{nazarov}
calculated the LDOS of a F film in contact with a superconductor. They
assumed that the electrons in the F film are scattered only at the rough
F/Vac surface (Vac stands for Vacuum).  However, not so much attention
has been payed to the more realistic case $h\tau \gg 1$.  Except for Ref. 
\cite{larkin}, where this limit was considered for very thin ferromagnetic
films, a study of the proximity effect in disordered ferromagnetic conductors
with strong exchange energies $h$ is still lacking. This limit is very
important when dealing with ferromagnetic layers made of transition metals,
as Fe or Ni, for which $h\lesssim 1$eV, and whose sizes are larger than the mean
free path $l$.

In the present paper, we analyze the DOS and the critical temperature $T_{c}$
of the superconducting transition in S/F structures for which the condition

\begin{equation}
h\tau \gg 1  \label{clean}
\end{equation}
is satisfied. Here $\tau $ is the elastic relaxation time due to impurity
scattering in the bulk.  According to condition (\ref{clean})  the exchange interaction, which is proportional to 
$h$ is much stronger than the interaction with impurities, which
is of the order of $\tau ^{-1}$. We will show that in this limit the decay
length of the superconducting condensate function induced in the ferromagnet
is of the order of the mean free path $l$. Therefore, despite its smallness,
it is very important to retain the impurity scattering term if the thickness
of the F layer is larger than $l$. On the other hand, the product $\tau T_{c}
$ may be smaller or larger than unity. We will see that in the limit (\ref
{clean}) thermodynamical quantities, such as DOS and $T_{c}$, differ
considerably from those in the dirty \cite
{radovic,buzdin1,aarts,fazio,lazar,buzdin} and purely ballistic limits \cite
{nazarov,minnesota}.

As it has been pointed out in Refs.\cite{bula,BVE_prb}, the Usadel equation
is not applicable  in the limit (\ref{clean}), and the more general
Eilenberger equation should be solved.  One can solve the latter
equation with the help of an expansion in the parameter $(h\tau )^{-1}$ for
both the strong 
 and weak proximity effect. In the case of a weak
proximity effect the solution can be obtained  even for an arbitrary
impurity concentration \cite{BVE_prb}, i.e. for an arbitrary value of the
parameter $h\tau $. It will be shown that in the  limit of large $h\tau 
$ the condensate function $f$ oscillates in space with a period $v_{F}/h$
and decays  on a length of the order of the mean free path $l$. If one
neglects the impurity scattering in the bulk, as was done in Refs. \cite
{nazarov,minnesota}, the results for DOS depend essentially on the boundary
conditions at the F/Vac interface. This is because in that case the
condensate function $f$ $\ $ is formed by interfering waves reflected
from the F/Vac boundary. Therefore, the thermodynamical properties depend
sensitively on whether these reflections are specular \cite{minnesota} or
diffusive \cite{nazarov}. 

In the present paper, we assume that the thickness of the ferromagnetic
layer $d_{F}$ is much greater than the mean free path $l$ and therefore
reflected waves are unimportant provided the bulk impurity scattering
dominates. We show that in the main approximation in the parameter $(h\tau
)^{-1}$, the LDOS in the ferromagnetic region in a F/S bilayer is not
affected by the proximity effect. On the other hand the amplitude of the
condensate function in the ferromagnet near the S/F interface is equal to
that in the superconductor S if the interface transmittance is high, and
decays away from the interface  at distances of the order $l$.  In
the limit (\ref{clean}), the mean free path $l$ is much larger than the
period of the oscillations ($\sim v_{F}/h$) and may be comparable with $d_{F}
$. Therefore, one can speak about long-range penetration of the
superconducting condensate into the ferromagnet. In the presence of two
superconductors (e.g. in a S/F/S sandwich) the situation  changes: a
correction to the DOS in F arises in the main approximation and its magnitude depends on the phase
difference between the superconductors and on the thickness $d_{F}$ of the F
layer vanishing in the limit $d_{F}\rightarrow \infty $. We also analyze a
variation of the critical temperature of the superconducting transition $%
T_{c}$ for a S/F bilayer for which the condition (\ref{clean}) is satisfied.
{\bf We find the dependence of $T_c$ on the thickness $d_S$ of the superconductor in the limit $d_F\gg l$ and compare it with the results obtained in the dirty limit}.

\section{Basic Equations}

 Physical quantities of interest can be computed using quasiclassical
retarded $\hat{g}^{R}$ and advanced $\hat{g}^{A}$  Green's
functions. In order to find the retarded Green's function $\hat{g}^{R\text{ 
}}$(for brevity we omit the index $R$)  one should solve the
Eilenberger equation  (the advanced Green's function can be found in the
same way) 
\begin{equation}
\mu v_{F}\partial _{x}\hat{g}-i(\epsilon +h)\left[ \hat{\tau}_{3},\hat{g}%
\right] -i\left[ \hat{\Delta},\hat{g}\right] +(1/2\tau )\left[ \left\langle 
\hat{g}\right\rangle ,\hat{g}\right] =0\;.  \label{eilenberger}
\end{equation}
Here $\hat{g}$ is the retarded Green function in the Nambu space. It obeys
the normalization condition 
\begin{equation}
\hat{g}^{2}=1\;.  \label{normalization}
\end{equation}
The angle brackets denote the averaging over angles: $\langle (...)\rangle
=\int_{0}^{1}{\rm d}\mu (...)$, $\mu =\cos \theta $ , $\theta $ is the angle
between the momentum and the $x$-axis (in all the S/F systems considered
below the $x$-axis is perpendicular to the S/F interfaces), and $v_{F}$ is
the Fermi velocity. The exchange field $h$ vanishes in the S-layers. $\hat{%
\Delta}$ is the matrix

\[
\hat{\Delta}=\left( 
\begin{array}{cc}
0 & \Delta  \\ 
-\Delta ^{\ast } & 0
\end{array}
\right) \;,
\]
and $\Delta $ is the pair potential in the superconductor. We assume that in
the F-layers the electron-electron interaction vanishes and therefore $%
\Delta =0$. The last term of Eq.(\ref{eilenberger}) describes the effect of
nonmagnetic impurities. Eq.(\ref{eilenberger}) is supplemented by the
Zaitsev boundary condition at the S/F interface \cite{zaitsev}: 
\begin{equation}
\hat{a}\left( R-R\hat{a}^{2}+\frac{T}{4}(\hat{s}_{1}-\hat{s}_{2})^{2}\right)
=\frac{T}{4}[\hat{s}_{2},\hat{s}_{1}]\;.  \label{zaitsev}
\end{equation}
Here $\hat{a}$ and $\hat{s}$ are the antisymmetric and the symmetric (in $%
\mu $) parts of the Green's function $\hat{g}$; $R$ and $T$ are the
reflection and transmition coefficients. Eqs. (\ref{eilenberger}-\ref
{zaitsev})  describe the system completely and are the basic equations 
 in our study.  In spite of a rather simple symbolic representation
they are quite complicated and a general solution can hardly be obtained for
an arbitrary impurity concentration. Accordingly, further calculations are
performed under certain assumptions that allow us to simplify  equations (\ref{eilenberger}-\ref{zaitsev}) for values of parameters which still
correspond to real systems.

\section{Local density of states in a S/F bilayer}

 In this section, it is assumed that the mean free path $l_{s}$ in the
superconductor is larger than the coherence length $\xi _{s}$. In this case
the last term in Eq. (\ref{eilenberger}) can be disregarded in the S region. We consider
first a S/F bilayer. The S/F interface is located at $x=0$, the
superconductor and ferromagnet occupy the regions $x<0$ and $x>0$,
respectively. We also assume that the thickness of  both layers are much
larger than the characteristic lengths over which $\hat{g}$ varies, {\it i.e}
$\xi _{s}$ in the S-layer and $l$ in the F layer. We focus first on the
LDOS in the superconductor for subgap energies: $\epsilon
<\Delta $. It is clear that in this energy range and for $|x|\gg \xi _{s}$
the LDOS vanishes as it should be in a bulk superconductor.

We assume a perfect S/F interface transparency. In this case and according
to the boundary condition Eq. (\ref{zaitsev}) both the symmetric and the
antisymmetric parts must be continuous at the interface. For simplicity we
approximate the dependence $\Delta (x)$ by a step like function $\Delta
(x)=\Delta \Theta (-x)$. Then, the solution of Eq. (\ref{eilenberger}) in
the superconducting region has the form 
\begin{equation}
\hat{g}_{S}=\left( \epsilon /\xi -{\rm sgn}\mu \cdot \Delta /\xi Ce^{\kappa
_{s}x}\right) \hat{\tau}_{3}+\left( \Delta /\xi -{\rm sgn}\mu \cdot
(\epsilon /\xi )Ce^{\kappa _{s}x}\right) i\hat{\tau}_{2}+C{\rm sgn}\mu e^{\kappa _{s}x}%
\hat{\tau}_{1}\;.  \label{greens1}
\end{equation}
Here $\xi =\sqrt{\epsilon ^{2}-\Delta ^{2}}$, $\kappa _{s}=-2i\xi /|\mu
|v_{F}$ and $C$ is a constant which has to be determined from the boundary
conditions.  Note that the solution  (\ref{greens1}) can also be
represented in terms of the eigenfunctions $\hat{U}=(u,v)$ of the Bogolyubov
equation, {\it i.e.} it can be written as 
\begin{equation}
\hat{g}_{S}=\hat{g}_{Sb}+\hat{\tau}_{3}\hat{U}^{+}(x)\otimes \hat{U}(x)\;,
\end{equation}
 where $\hat{g}_{sb}$ is the Green function in the bulk. For energies $%
|\epsilon |<\Delta $ the functions $\hat{U}(x)$ decay exponentially away
from the S/F interface. 

Now we have to solve Eq. (\ref{eilenberger}) in the ferromagnet. In the
limit determined by Eq. (\ref{clean}) one can check that the solution of Eq. (\ref
{eilenberger}) in the main approximation with respect to $(h\tau )^{-1}$ is 
\begin{equation}
\hat{g}_{F}=\hat{\tau}_{3}+D\exp (-\kappa x/|\mu |l)i\hat{\tau}_{2}+{\rm sgn}%
\mu \cdot D\exp (-\kappa x/|\mu |l)\hat{\tau}_{1}\;,  \label{greenf1}
\end{equation}
where $\kappa =1-2i(\epsilon +h)\tau $.  The second and third terms in
Eq. (\ref{greenf1}) describe the condensate function induced in the
ferromagnet due to the proximity effect. 

\begin{figure}
\epsfysize = 10cm
\centerline{\epsfbox{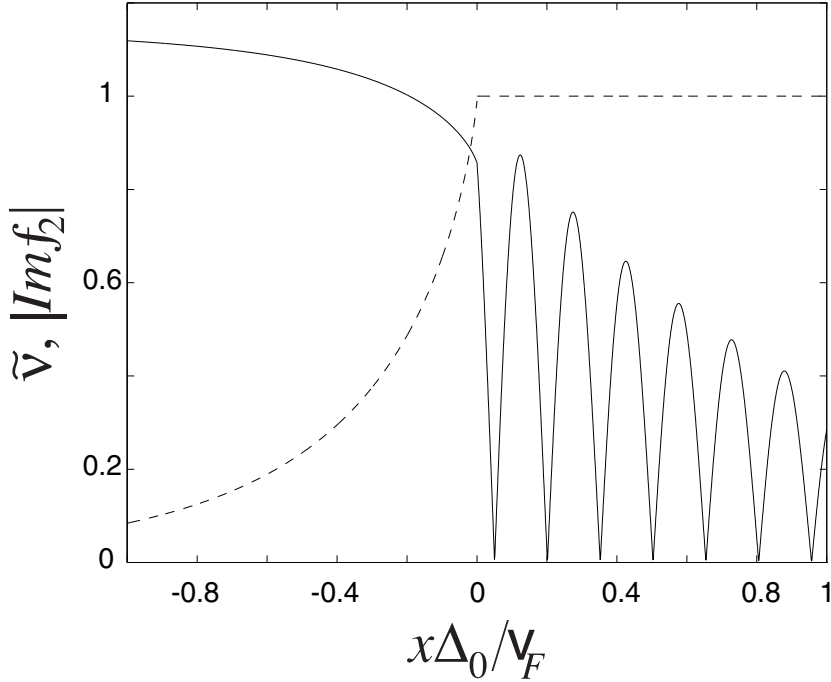
 }}
\caption{The spatial dependence of the normalized LDOS $\tilde{\nu}$ (dashed line), and the imaginary part
  of the symmetric condensate function $|{\rm Im}f_2|$ (solid line). The S (F) layer is located in
the region $x<0$ ($x>0$). $\Delta _{0}$ is the value of $\Delta $ for $T=0$. 
$\Delta /\Delta _{0}=0.8$, $\protect\epsilon /\Delta
_{0}=0.4$,$\tau\Delta_0=1$ and $h/\Delta_0=10$. Although the 
function $|{\rm Im}f_2|$ exhibits a strongly oscillating and weakly decaying behavior in the ferromagnet,
the LDOS remains unchanged in this region.}
\label{Fig.1}
\end{figure}

 As opposed to many authors claiming that the decay length of the
condensate function is proportional to $v_{F}/h$ (see for example Ref.\cite
{aarts}), Eq. (\ref{greenf1}) clearly shows that the penetration depth of
the condensate is of the order of the mean free path $l$ and does not depend
on the exchange field $h$.  Thus, for clean and strong ferromagnets, the superconducting condensate penetrates  over long distances compared to the magnetic length $v_F/h$. As mentioned in the introduction, in
the dirty case, $h\tau \ll 1$, the penetration length of the condensate into
the ferromagnet decreases with increasing $h$, which is true for the singlet
component of the condensate function (the triplet component of the
condensate may penetrate into the ferromagnet over a long length of the
order $\sqrt{D/T}$ \cite{BVE_prl,kadigrobov}. This triplet component may be
induced by the presence of a magnetic inhomogeneity close to the S/F
interface\cite{BVE_prl}). However, we see from Eq. (\ref{greenf1}) that  a
long-range proximity effect exists even for singlet pairing and homogeneous
magnetization. Using the fact that $\hat{g}$ is continuous at $x=0,$ one
can determine the constants $D$ and $C$ from Eqs. (\ref{greens1}) and (\ref
{greenf1}) 
\begin{equation}
C=D=\frac{-\xi+\epsilon }{\Delta }  \label{D_C}
\end{equation}
The normalized LDOS is given by the well known expression 
\begin{equation}
\tilde{\nu}=\nu /\nu _{0}=(1/2){\rm ReTr}\hat{\tau}_{3}\hat{g}\;,
\label{dos}
\end{equation}
where $\nu _{0}$ is the LDOS in the normal state.  From Eqs. (\ref
{greens1}-\ref{dos}) one obtains in the S region ($\epsilon<\Delta$) 
\begin{equation}
\tilde{\nu}(x,\epsilon )=<\exp (2\sqrt{\Delta ^{2}-\epsilon ^{2}}x/|\mu
|v_{F})>\;.  \label{result1}
\end{equation}
In Fig.\ref{Fig.1}, we plot the spatial dependence of $\tilde{\nu}$ and the
absolute value of the imaginary part of the symmetric condensate function
(second term in Eq. (\ref{greenf1})). Due to the proximity effect of the
ferromagnet the superconductor becomes gapless in the region close to the
S/F interface.  According to Eqs. (\ref{greenf1}) and (\ref{dos}) the
LDOS in the ferromagnet remains unchanged despite the presence of the
condensate function induced in the F region (second and third term in Eq. (%
\ref{greenf1})). The reason for that is the cancellation of the symmetric
(second term in Eq. (\ref{greenf1})) and antisymmetric parts (third term in
Eq. (\ref{greenf1})) in the normalization condition (\ref{normalization}).
Thus, the coefficient in front of $\tau _{3}$ in Eq. (\ref
{greenf1}) is 1 and therefore $\tilde{\nu}=1$. 

\begin{figure}[tbp]
\epsfysize = 7cm
\vspace{0.2cm}
\centerline{\epsfbox{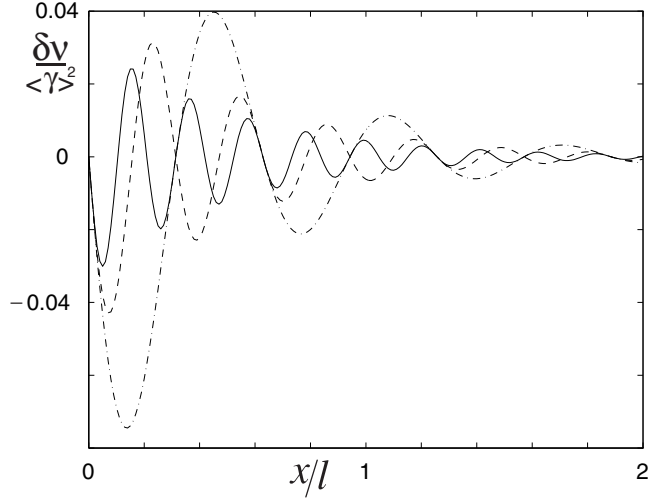
 }}
\vspace{0.2cm}
\caption{Spatial dependence of the LDOS in the ferromagnet for different
values of $h\protect\tau$. The solid, dashed and dash-dotted lines correspond
to $h\protect\tau=15$, $h\protect\tau=10$ and  $h\protect\tau=5$, respectively.}
\label{Fig.2}
\end{figure}

The correction to the LDOS  in the ferromagnet is not zero if one takes
into account higher order terms in the expansion in the parameter $(h\tau
)^{-1}$. We calculate this correction assuming for simplicity that the
transparency of the S/F interface is low. In this case, the Green function
in the superconductor is not affected by the proximity effect and is given
by 
\begin{equation}
\hat{g}_{S}=G_{S}\hat{\tau}_{3}+F_{S}i\hat{\tau}_{2}\;,  \label{greens2}
\end{equation}
where $G_{S}=\epsilon /\sqrt{\epsilon ^{2}-\Delta ^{2}}$ and $F_{S}=\Delta /%
\sqrt{\epsilon ^{2}-\Delta ^{2}}$. For a low interface transparency, the
boundary condition, Eq. (\ref{zaitsev}), reduces to 
\begin{equation}
\hat{a}=\gamma \left[ \hat{s}_{F},\hat{s}_{S}\right] \cong \gamma F_{S}\hat{%
\tau}_{1}\;.  \label{lowtrans}
\end{equation}
Here $\gamma =T(\mu )/4R(\mu )\ll 1$ is the parameter describing the
transmittance of the interface. The way how to solve Eq. (\ref{eilenberger})
was presented in Ref. \cite{BVE_prb}. The exact Green function is given by

\begin{equation}
\hat{g}_{F}(x)=\hat{\tau}_{3}+f_{2}(x)i\hat{\tau}_{2}+f_{1}(x)\hat{\tau}_{1}=%
\hat{\tau}_{3}+\int ({\rm d}k/2\pi )(f_{2k}i\hat{\tau}_{2}-(\mu l/\kappa
)f_{2k}\hat{\tau}_{1}\partial _{x})e^{ikx}\;,
\end{equation}
where 
\begin{equation}
f_{2k}=\frac{2\kappa F_{S}l}{M\left[ 1-\kappa \langle 1/M\rangle \right] }%
\left[- \kappa \left( \gamma \mu \langle 1/M\rangle -\langle \gamma \mu
/M\rangle \right) +\gamma \mu \right] \;,  \label{fourier}
\end{equation}
$M=(kl\mu )^{2}+\kappa ^{2}$ and $\kappa =1-2i(\epsilon +h)\tau $. In the
case under consideration, {\it i.e} $h\tau \gg 1$, one can easily show that $%
|\kappa \langle 1/M\rangle| <1$ and hence we can expand Eq. (\ref{fourier}): 
\begin{equation}
f_{2k}\cong \frac{2\kappa F_{S}l}{M}\left( \gamma \mu +\kappa \gamma
_{0}/M_{\mu =1}\right) \;.  \label{expansion}
\end{equation}
For simplicity we have assumed that the transmission coefficient $T(\mu )$
has a sharp maximum at $\theta =0,$ i.e. $\gamma (\mu )=\gamma _{0}\delta
(\mu -1)$. The first correction to the normalized LDOS proportional to $%
(h\tau )^{-1}$ can be now obtained. After cumbersome but straightforward
calculations one obtains at low energies 
\begin{equation}
\delta \nu =\frac{1}{2}{\rm Re}\langle (f_{2})^{2}-(f_{1})^{2}\rangle
=-\gamma _{0}^{2}(e^{-2x/l}/4h\tau )\sin (4h\tau x/l)\;.  \label{doscorr}
\end{equation}
 Eq. (\ref{doscorr}) shows that for sufficiently large $h\tau ,$ the
correction to the LDOS  of the normal metal is small. This
correction to the LDOS proportional to $(h\tau )^{-1}$  reveals a
damped-oscillatory behavior similar as the one reported in Ref.\cite{buzdin}%
, where the case $h\tau \ll 1$ (dirty limit) was considered. The spatial
dependence of the local DOS is shown in Fig.\ref{Fig.2}.


\section{Local density of states in a S/F/S sandwich}

\begin{figure}
\epsfysize = 7cm
\centerline{\epsfbox{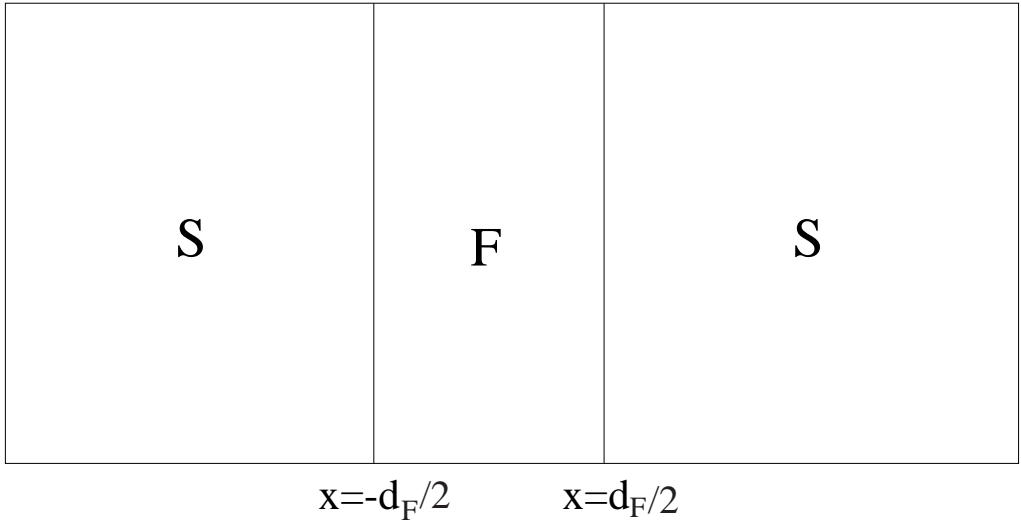
 }}
\caption{The S/F/S structure.}
\label{Fig.3}
\end{figure}

Now we consider a Josephson-like S/F/S structure (see Fig.\ref{Fig.3}). In
this case, the correction to the DOS is not zero even in the main
approximation in the parameter $(h\tau )^{-1}$. The thickness of the F-layer
is $d_{F}$, and we assume again that the transparency of the S/F interfaces
is low. The solution of Eq. (\ref{eilenberger}) can be  sought in the
following form 
\begin{equation}
\hat{g}=\hat{\tau}_{3}+\hat{s}+\hat{a}\;,  \label{greenf3}
\end{equation}
where $\hat{s}$ and $\hat{a}$ are the symmetric and antisymmetric part of
the condensate function induced in the ferromagnet. The equations which
determine $\hat{s}$ and $\hat{a}$ can be obtained from Eq. (\ref{eilenberger}%
)  and written in the form 
\begin{equation}
\begin{array}{c}
(\mu l)^{2}\partial _{xx}^{2}\hat{s}-\kappa ^{2}\hat{s}-\kappa \langle 
\hat{s}\rangle =0 \\ 
\hat{a}=-(\mu l/\kappa )\hat{\tau}_{3}\partial_x\hat{s}\; ,
\end{array}
\label{eil_dec}
\end{equation}
where $\kappa$ is defined in Eq. (\ref{fourier}).

\begin{figure}
\epsfysize = 7cm
\vspace{0.2cm}
\centerline{\epsfbox{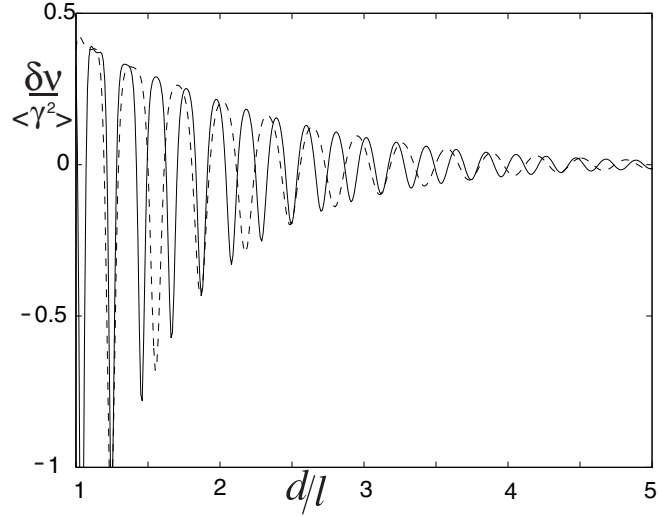
 }}
\vspace{0.2cm}
\caption{LDOS  versus the thickness $d_F$ of the F-layer. Here $%
\protect\epsilon\protect\tau=0.1$, $\Delta\protect\tau=1$ and $\protect%
\varphi=0$. The solid and dashed lines correspond to $h\protect\tau=15$ and $h\protect\tau=10$.}
\label{Fig.4}
\end{figure}

\begin{figure}
\epsfysize = 6cm
\vspace{0.2cm}
\centerline{\epsfbox{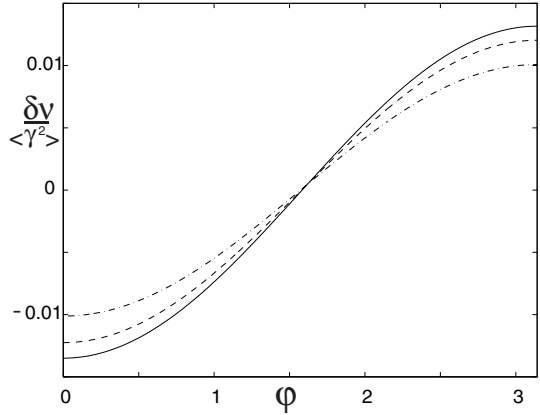
 }}
\vspace{0.2cm}
\caption{LDOS  versus the phase difference $\protect\varphi $ for the
parameters $\protect\epsilon \protect\tau =0.1$, $\Delta \protect\tau =1$
and $d_F/l=5$. The solid, dashed and dash-dotted lines correspond to
$h\protect\tau =15$,  $h\protect\tau =10$ and  $h\protect\tau =5$, respectively.}
\label{Fig.5}
\end{figure}

In the limit of a weak proximity effect, the boundary conditions at $x=\pm
d_{F}/2$ are [the upper (lower) sign corresponds to $x=d_{F}/2$ ($x=-d_{F}/2$%
)] 
\begin{equation}
\hat{a}=\mp \gamma \hat{\tau _{3}}\hat{F_{S}}\;.  \label{boundcond3}
\end{equation}
$\hat{F_{S}}$ is the condensate function of the superconductors which now is
given by $\hat{F_{S}}(\varphi )=i\hat{\tau}_{2}F_{S}\exp (\pm i\varphi \hat{%
\tau}_{3}),$ where $F_{S}$ is defined in Eq.(\ref{greens2}) and $\varphi $ is the phase
difference between the superconductors. The general solution of Eq. (\ref
{eil_dec}) with the boundary conditions Eq.(\ref{boundcond3}) was presented
in Ref. \cite{BVE_prb}. In the  limit $h\tau \gg 1$ one obtains 
\begin{equation}
\begin{array}{ccl}
\hat{s}(x) & = & -2\kappa _{F}l\mu \gamma F_{S}\sum\limits_{n}\left( i\hat{%
\tau}_{2}\cos \varphi /2+i\hat{\tau}_{1}(-1)^{n}\sin \varphi /2\right) \int 
\frac{{\rm d}k}{2\pi }e^{-ikx}e^{ik(2n+1)d_F}/M \\ 
& = & -\gamma F_{S}\left[ i\hat{\tau}_{2}\cos (\varphi /2)\frac{\cosh \theta
(x)}{\sinh \theta (d_{F}/2)}+i\hat{\tau}_{1}\sin (\varphi /2)\frac{\sinh
\theta (x)}{\cosh \theta (d_{F}/2)}\right] \;,
\end{array}
\label{solution3}
\end{equation}
where $\theta (x)=\kappa x/l|\mu |$ and $n$ takes integer values from $%
-\infty $ to $+\infty $. From Eq. (\ref{eil_dec}) one obtains the expression
for $\hat{a}$ 
\begin{equation}
\hat{a}(x)=\gamma F_{S}\left[ \hat{\tau}_{2}\sin (\varphi /2)\frac{\cosh
\theta (x)}{\cosh \theta (d_{F}/2)}-\hat{\tau}_{1}\cos (\varphi /2)\frac{%
\sinh \theta (x)}{\sinh \theta (d_{F}/2)}\right] \;.  \label{solution3a}
\end{equation}
The correction to the LDOS due to the proximity effect is then given by 
\begin{equation}
\delta \nu =-\frac{1}{2}{\rm Re}\left( \hat{s}^{2}+\hat{a}^{2}\right) =2%
\frac{\Delta ^{2}}{\epsilon ^{2}-\Delta ^{2}}{\rm Re}\langle \frac{\gamma
^{2}}{\sinh ^{2}2\theta (d_{F}/2)}\left( 1+\cos \varphi \cosh 2\theta
(d_{F}/2)\right) \rangle \; ,  \label{dos3}
\end{equation}
where $2\theta(d_F/2)=\kappa d_F/l|\mu|$
As  one could expect, $\delta \nu $ vanishes in the limit
of a very large thickness $d_{F}$  of the F-layer. At the same time,
one can see that the correction to the LDOS is not x-dependent but it
depends on the phase difference $\varphi $. This behavior is quite
interesting and rather unexpected. The correction to the LDOS, Eq. (\ref
{dos3}), may be both positive and negative depending on the phase $\varphi $%
. In Fig.\ref{Fig.4} and Fig.\ref{Fig.5} we plot the dependence of $\delta
\nu $ on  the thickness $d_{F}$ and on the phase difference $\varphi $. {\bf As before we assume that $\gamma^2 (\mu )=(\gamma^2 _{0}/\delta\mu)\delta (\mu
-1)$, where $\delta\mu$ is the width of the peak in the dependence $\gamma(\mu)$. Thus, for a S/F/S sandwich the  correction to the LDOS $\delta\nu$ is much larger than in the case of a S/F bilayer. In the latter case the  correction is proportional to the small factor $(h\tau)^{-1}$, while in the S/F/S structure $\delta\nu$ is finite in zeroth order in $(h\tau)^{-1}$ and proportional to the large factor $1/\delta\mu$.}


\section{Critical Temperature for a S/F bilayer}

Another thermodynamic quantity we are interested in, is the superconducting
critical temperature of a S/F bilayer in the limit $h\tau \gg 1$. The
opposite case ($h\tau \ll 1$) was studied in Refs.\cite
{radovic,buzdin1,aarts,lazar}. Here we consider a bilayer system consisting
of one S-layer of thickness $d_{S}$ and one F-layer of thickness $d_{F}$. 
 Of course, the superconducting transition temperature can considerably
change only provided the thickness of the superconducting layer is not very
large $d_{s}\lesssim \xi _{S}.$ We assume that the S/F interface has a very
high transparency,  otherwise the proximity effect does not affect
substantially the critical temperature $T_{c}$. The  computation of
the condensate function $\hat{f}=\hat{s}+\hat{a}$ from Eq. (\ref{eilenberger}%
) follows as before. In the limit $h\tau \gg 1$ the solution in the F-layer
is given  simply by Eq. (\ref{greenf1}). The solution in the S region
requires a little more care. For temperatures close to $T_{c}$ one can seek
for solutions in the form 
\begin{equation}
\hat{g}_{S}=\hat{\tau}_{3}+F_{2}i\hat{\tau}_{2}+F_{1}\hat{\tau}_{1}\;.
\label{ansatzS}
\end{equation}
If we assume that $\Delta (x)$ is a slow varying function of $x$, then the
function $F_{2}$ can be written as \cite{foot} 
\begin{equation}
F_{2}=\Delta /\epsilon +\delta F\;  \label{ansatzF}
\end{equation}
Bearing in mind that the functions $F_{1}$ and $F_{2}$ are continuous at the
S/F interface, one obtains the following equation for $\delta F$\cite{foot2}
\begin{equation}
(\mu l)^{2}\partial _{xx}^{2}\delta F-\kappa _{s}^{2}\delta F-\kappa
_{s}\langle \delta F\rangle =-2|\mu |l\kappa _{s}D\delta (x)\;,
\label{equationf2}
\end{equation}
where $\kappa _{s}=-1+2i\epsilon \tau $, and D is the constant from Eq. (\ref
{greenf1}). Equation (\ref{equationf2}) can be easily solved in the Fourier
space. The solution is given by 
\begin{equation}
\delta F_{k}=\frac{2l\kappa _{s}D}{M_{S}\left( 1+\kappa _{s}\langle
1/M_{S}\rangle \right) }\left( |\mu| +|\mu| \kappa _{s}\langle 1/M_{S}\rangle
-\kappa _{s}\langle |\mu| /M_{S}\rangle \right) \;.  \label{fourier4}
\end{equation}
Here $M_{S}=(kl\mu )^{2}+\kappa _{s}^{2}$. In the dirty limit. {\it i.e.}
when $\epsilon \tau \ll 1$, the expression Eq. (\ref{fourier4}) can be
simplified 
\begin{equation}
\delta F_{k}\cong \frac{3l\kappa _{s}D}{(lk)^{2}-6i\epsilon \tau }\;.
\label{dirtyf}
\end{equation}
and the value of the function $\delta F(x)$ at $x=0$ is given by 
\begin{equation}
\delta F(0)=\int \frac{{\rm d}k}{2\pi }\delta F_{k}=-\frac{\sqrt{3}(1+i)}{4%
}\frac{D}{\sqrt{\epsilon \tau }}\;.  \label{valuex0}
\end{equation}

From Eqs. (\ref{greenf1}), (\ref{ansatzF})-(\ref{valuex0}) and the fact that
all the functions are continuous at $x=0$, one can determine the constant $D$ 
\begin{equation}
D=\frac{\Delta}{\epsilon} \frac{4\sqrt{\epsilon \tau }}{\sqrt{3}(1+i)}\;.  \label{C}
\end{equation}
We see that the constant $D$ does not depend on the strength of the exchange
field $h$, and therefore the condensate function in the superconductor is
not $h$-dependent. Thus,  we come to {\bf the following} conclusion: in the
limit $h\tau \gg 1$ and if $d_F>l$  the thermodynamic  quantities of the superconductor
do not depend on the strength of the exchange field $h$; in particular, the
critical temperature of the bilayer does not depend on the thickness of the
ferromagnetic layer.  This result is in qualitatively agreement with experimental data
presented  in Ref.\cite{strunk}, in which the
dependence  $T_c^*(d_{Gd})$ of Nb/Gd systems  was determined and a saturation of $T_c^*$ for large $d_{Gd}$ was observed. 

In order to estimate the critical temperature $T_{c}^{\ast }$ of the system,
we need to calculate the pair potential $\Delta (x)$. The latter satisfies
the self-consistency equation 
\begin{equation}
\Delta =\lambda \int {\rm d}\epsilon \tanh (\epsilon \beta
)(F_{2}^{R}-F_{2}^{A})  \label{self}
\end{equation}
We notice that all equations presented above are for the retarded component.
The advanced Green function $F_{2}^{A}$ can be obtained using the  
relation $(F_{2}^{R})^{\ast }=-F_{2}^{A}$. One can easily check that at $x=0
$ the integral in Eq. (\ref{self}) diverges as $1/\sqrt{\epsilon }$ and
therefore the function $\Delta (x)$ must vanish at $x=0$. Thus, the
boundary condition for the pair potential at the S/F interface can be
written as 
\begin{equation}
\Delta (0)=0\;.  \label{bc_delta}
\end{equation}
Using the Ginsburg-Landau equation
\begin{equation}
\xi _{GL}^{2}\Delta ^{^{\prime \prime }}=-\Delta \;,  \label{LG}
\end{equation}
where $\xi _{GL}=\xi _{0}/\sqrt{1-T_{c}^{\ast }/T_{c}}$, and the boundary
condition Eq. (\ref{bc_delta}) we obtain 
\begin{equation}
\Delta =A\sin (x+d_s)/\xi _{GL}\;.
\end{equation}
At the boundary with the vacuum ($x=-d_{s}$) $\partial _{x}\Delta =0$. From
this condition we obtain an expression which determines $T_{c}^{\ast }$ 
\begin{equation}
1-T_{c}^{\ast }/T_{c}=\left( \pi \xi _{0}/2d_{s}\right) ^{2}\;.  \label{tc}
\end{equation}
We remind that equation (\ref{tc}) is only valid in the case $|T_{c}^{\ast
}-T_{c}|<T_{c}$. {\bf It gives the asymptotic value of $T_c^*$ in the limit $h\tau\gg1$ and $d_F\gg l$.}  This formula describes the usual proximity effect in a S/N structure: $%
T_{c}^{\ast }$ coincides with $T_{c}$ if $d_{s}>>\xi _{0}$ and decreases
with increasing $d_{s}$ Ref. \cite{deGennes}. {\bf The same expression
for $T_c^*$ is valid in the dirty limit ($h\tau\ll1$, see Ref. \cite{aarts}).}
\section{Conclusion}

In conclusion, we have considered different S/F structures in the case of a
strong exchange field $h$,  when the condition $h\tau \gg 1$ is
satisfied. The results of our analysis differ considerably from those
obtained in the ``dirty''\cite{radovic,buzdin1,aarts,fazio,lazar,buzdin} and
in the pure ballistic limit\cite{minnesota}. We have shown that the
thermodynamic properties ( such as local DOS and critical temperature)
of a S/F bilayer do not depend on $h$ in the main approximation with respect
to the small parameter $(h\tau )^{-1}$.  It is worth mentioning that in
this approximation the LDOS does not change in the ferromagnet below $T_{c}$%
. The superconducting gap is however suppressed near the S/F interface
(gapless state) and restored at distances of the order of $\xi _{s}\sim
v_{F}/\Delta $. In a S/F/S sandwich the correction to the LDOS in the
ferromagnet is nonzero, spatially constant, and depends on the phase
difference $\varphi $ between the superconductors. The different behavior of
the LDOS in a S/F and S/F/S structure is due to the interference of the
induced condensate functions created at each S/F interface. Thus, the
changes of the LDOS $\nu $ due to the proximity effect may be observed more
clearly in a S/F/S structure by measuring the dependence of $\nu $ on the
phase difference $\varphi $ between the superconductors. 

We would like to thank SFB 491 {\it Magnetische Heterostrukturen} for
financial support.


\begin{references}
\bibitem{foot}  This assumption is justified when $|T_c^*-T_c|<T_c$, where $%
T_c^*$ is the transition temperature of the bilayer and $T_c$ the critical
temperature of the bulk superconductor.
\bibitem{foot2} We continue $\delta F$ for positive $x$ in a symmetric way;
  however the solution we need corresponds to negative values of $x$.
\end{references}
\end{document}